\documentclass[
 amsmath,amssymb,
 aps,
 nofootinbib,
pra,reprint
]{revtex4-1}

\usepackage{graphicx}
\usepackage{dcolumn}
\usepackage{bm}
\usepackage{units}
\usepackage{siunitx}
\graphicspath{{./pictures/}} 
\usepackage[]{amsmath,amsfonts,amssymb}
\usepackage{physics}
\usepackage{color}
\usepackage[symbol*]{footmisc}

\DefineFNsymbolsTM{otherfnsymbols}{%
a \circ
b  \mathsection}

\setfnsymbol{otherfnsymbols}

\usepackage{natbib}
\setcitestyle{sort&compress,numbers,square}

\begin{document}

\title{Axions as a probe of solar metals}

\author{Joerg Jaeckel }
 \email{\href{mailto: jjaeckel@thphys.uni-heidelberg.de}{jjaeckel@thphys.uni-heidelberg.de}}
\author{Lennert J. Thormaehlen}%
 \email{\href{mailto: l.Thormaehlen@ThPhys.Uni-Heidelberg.DE}{l.thormaehlen@thphys.uni-heidelberg.de}}
\affiliation{%
Institut f\"ur Theoretische Physik, Universit\"at Heidelberg, Philosophenweg 16, 69120 Heidelberg, Germany
}%

\begin{abstract}
Axion helioscopes aim to detect axions which are produced in the core of the sun. Their spectrum contains information about the solar interior and could in principle help to solve the conflict between high and low metallicity solar models. Using the planned \textit{International Axion Observatory} as an example, we show that helioscopes could measure the strength of characteristic emission peaks caused by the presence of heavier elements with good precision. In order to determine unambiguously the elemental abundances, an improved modelling of the states of atoms inside the solar plasma is required.

\end{abstract}

\maketitle

\section{Introduction}
\vspace{-0.1cm}
\subsection{The solar abundance problem}
\vspace{-0.1cm}
The sun has been the object of intensive studies and by now a detailed picture of processes inside the sun has emerged \cite{Bahcall:1987jc,Bahcall:2000nu}. Especially the important parameters pressure, temperature, density, hydrogen mass fraction and radiative opacities are well described by the so-called standard solar models \cite{Bahcall:2000nu}, which have proved very reliable even though they were questioned at the time when the solar neutrino problem had not yet been solved \cite{Bahcall:1996qw}. However, details of the chemical composition of the sun remain under discussion with conflicting measurements being made, resulting in the solar metallicity\footnote{According to astrophysical convention ``metals'' here refers to all elements heavier than helium.} or solar composition problem \cite{Bahcall:2004yr,Antia:2005mg,PenaGaray:2008qe,Asplund:2009fu,Serenelli:2009yc}.

The abundance of heavy elements inside the sun (metallicity) is quantified by $Z/X$, where $Z$ is the mass fraction of elements heavier than helium and $X$ the mass fraction of hydrogen. Abundances of metals can be estimated in several ways (see \cite{Asplund:2009fu} for a detailed overview). Helioseismic measurements \cite{Bahcall:2000nu,ChristensenDalsgaard:2002ur,Antia:2006zy,Basu:2007fp} consistently prefer high-$Z$ models ($Z/X=0.0245$ \cite{Bahcall:2000nu}) while modern photospheric measurements\footnote{All results since the major revision of \cite{Grevesse:1998bj} by Asplund et al. \cite{Asplund:2004eu}.} reach a significantly lower $Z$ ($Z/X=0.0178$ \cite{Asplund:2009fu}). This discrepancy was even larger~\cite{Asplund:2004eu} before the photospheric model AGSS09 was published \cite{Asplund:2009fu}, which revised $Z$ upwards\footnote{The most recent solar abundances of heavy elements from spectroscopic observations can be found in~\cite{Scott:2014lka,Scott:2014mka,Grevesse:2014nka}. They do not change the general picture \cite{Serenelli:2016dgz}.}. Photospheric models also predict the base of the solar convective envelope and the surface helium mass fraction which disagree with helioseismic data at 5 and \SI{11}{\sigma}, respectively \cite{Serenelli:2009yc}.

This constitutes an important problem in solar modelling and so far no satisfactory solution has been found, even though alternative solar models as well as new metallicity measurements have been considered and the required opacity calculations have been updated \cite{Antia:2005mg,Asplund:2009fu,2009ASPC..416..193B,Vinyoles:2016djt,Vagnozzi:2016cmr,Vagnozzi:2017wge}. 

\subsection{Solar axions}
Axions are light, weakly interacting pseudoscalar particles originally proposed to solve the strong CP problem~\cite{Peccei:1977hh, Peccei:1977ur,Weinberg:1977ma,Wilczek:1977pj}. Axions as well as their relatives, axion-like particles\footnote{Usually, the term axion-like particles refers particles similar to the QCD axion in that they are light (pseudo-)scalars and have very weak couplings to Standard Model particles, but they do not solve the strong CP problem. For our purposes only the two-photon and electron couplings are relevant. Their mass can be different from that of the QCD axion and can therefore be treated as a free parameter.}(cf. e.g.~\cite{Jaeckel:2010ni,Marsh:2015xka} for reviews), have since become attractive dark matter candidates~\cite{Preskill:1982cy,Abbott:1982af,Dine:1982ah,Arias:2012az} and may also be able to explain astrophysical observations such as the anomalous cooling of stellar objects and the gamma-ray transparency of the universe (cf.~\cite{Armengaud:2019uso} for a comprehensive overview). For simplicity we will henceforth just talk about axions, but axion-like particles are meant to be included.

If axions exist, they would be emitted in large numbers by the sun and could be detected on earth with an axion helioscope \cite{Sikivie:1983ip}. Such an experiment consists of a strong magnetic field of strength $B$ and length $L$. If it is aimed towards the sun, axions can convert to X-ray photons by coupling to the magnetic field. The conversion probability of a light axion ($m_a\lesssim \SI{10}{\milli \electronvolt}$) to a photon is~\cite{Sikivie:1983ip}
\begin{equation}
P_{a \rightarrow \gamma}=\frac{g_{a\gamma}^2 B^2 L^2}{4}, \label{conversionmassless}
\end{equation}
where $g_{a\gamma}$ quantifies the strength of the coupling to photons. The X-rays can be focused and detected with an energy dependent efficiency $Q(\omega)$. The detected spectral flux of photons $\dv{\Phi_\gamma}{\omega}$ can therefore be related to the solar axion flux $\dv{\Phi_a}{\omega}$ via 
\begin{equation}
	\dv{\Phi_\gamma}{\omega}=Q(\omega)P_{a \rightarrow \gamma}\dv{\Phi_a}{\omega}. \label{gammaflux}
\end{equation}
The proposed helioscope IAXO~\cite{Irastorza:2011gs,Irastorza:2013dav,Armengaud:2014gea} will be able to improve the sensitivity in parameter space by more than one order of magnitude in comparison to the CAST experiment. This will enable new applications beyond a discovery of axions, for example the possibilities to measure the mass of axions as well as to measure both the axion-photon and the axion-electron coupling~\cite{Jaeckel:2018mbn,Dafni:2018tvj}. 
Beyond that, as we will argue in this letter, it can also serve as a new tool to measure astrophysical parameters\footnote{In a similar way, the discovery of axions in a haloscope~\cite{Sikivie:1983ip} experiment could allow to determine detailed properties of the local dark matter such as the velocity distribution inside the galaxy~\cite{Irastorza:2012jq,Jaeckel:2013sqa,Jaeckel:2015kea,OHare:2017yze,Knirck:2018knd}.}. 

In particular, we show that in a viable range of axion parameter space we can measure the strength of characteristic peaks in the emission spectrum of axions. These peaks are due to the atomic transitions of metals and are therefore directly linked to the abundance of these elements in the interior of the sun. However, their strength also depends on the plasma properties such as the occupation numbers of the relevant excited states. Comparing four different models, we find sizeable differences, preventing an unambiguous determination of metal abundances. Nevertheless, measurements of the peak strength would still give us valuable information. If modelling of the atomic states can be sufficiently improved, we could determine the elemental abundances to a level relevant for the solar abundance problem. Or, coming from the other direction, the peaks themselves can tell us about the atomic states in the environment of the sun's interior.

\section{Abundance measurements with axions}
\subsection{Metals and and the solar axion flux}
Particles produced in the core of the sun and leaving without further interactions can provide information on metal abundances without the need to consider stellar envelope effects. Indeed, neutrinos mostly from the CNO cycle have already been proposed as such a probe~\cite{PenaGaray:2008qe,Cerdeno:2017xxl}. However, since they are produced in nuclear processes, they can only provide information on elements involved in such processes therefore giving only limited sensitivity to heavier elements.

Axions are also produced in the core of the sun. Importantly, metals inside the sun lead to characteristic peaks in the part of the solar axion flux arising from a coupling to electrons. The peaks are due to bound-bound transitions of electrons and therefore not dependent on nuclear processes.

\begin{figure}
	\centering
	\includegraphics[width=0.96\linewidth]{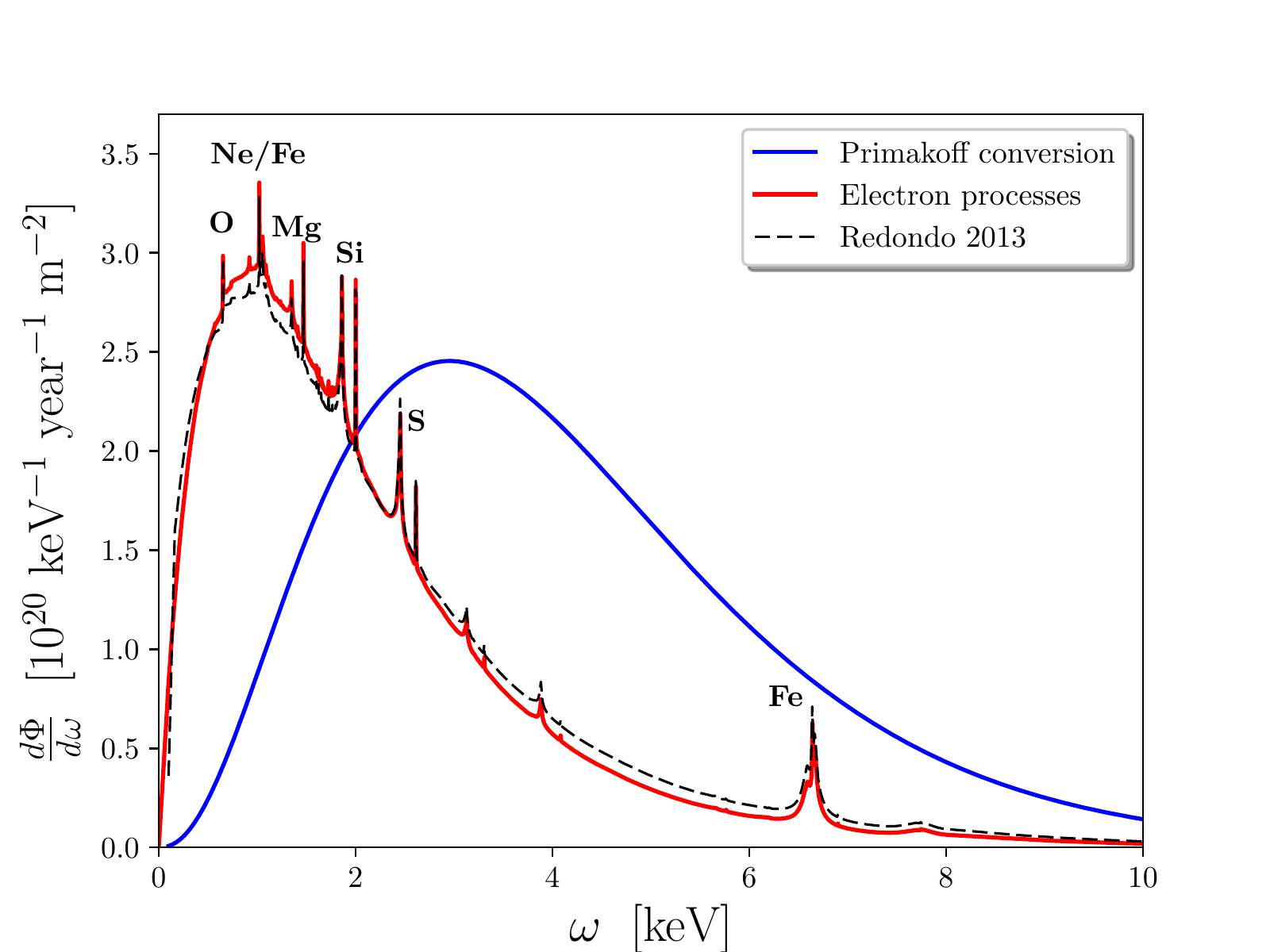}
	\caption{The solar axion flux from axion-electron processes (red) and Primakoff conversion (blue). Redondo's result~\cite{Redondo:2013wwa} (black dashed) was reproduced to good accuracy. Each pair of relevant Lyman lines is labelled with the element responsible. In the case of neon, these lines merge with the Balmer series of iron. It becomes clear from the plot that iron will be the easiest element to detect due to both a strong peak as well as a low background. We also plot the Primakoff contribution for $g_{a\gamma}=\SI{e-11}{\per \giga \electronvolt }$. If the Primakoff contribution dominates the background, elements like oxygen, whose peaks sit at a lower Primakoff flux, will be easier to detect.}
	\label{peak_labels}
\end{figure}

Following the derivation by Redondo in~\cite{Redondo:2013wwa}, the spectral axion emission can be related to monochromatic opacities provided by the Opacity Project (OP)~\cite{Badnell:2004rz,Seaton:2004uz} as well as OPAS~\cite{OPAS1,OPAS2}, ATOMIC~\cite{Colgan:2016} and LEDCOP~\cite{Magee:1995}. These have to be interpolated, as they are only available on a rough grid of plasma temperatures and electron densities. Using OP data and integrating the emission rates over the solar model, Redondo's result was recovered to good precision\footnote{\label{foot:three}We note a factor of 2 discrepancy in the Compton contribution. The plotted result in~\cite{Redondo:2013wwa} being larger than the one given in the equation Eq.~(2.19), which gives the correct result. We are indebted to Javier Redondo for clarifying this. Our computation also features a slightly higher contribution from free-free transition at low energies (most likely due to some difference in the approximation of the Debye screening scale). Since both contributions are continuous and the Compton contribution is by far the smallest one, this small discrepancy does not significantly impact our results.}, as can be seen in Fig.~\ref{peak_labels}.

For our calculations, we use the low-$Z$ photospheric model AGSS09~\cite{Serenelli:2009yc}.
Fig.~\ref{peak_labels} identifies the elements responsible for each significant peak. These are the ones whose abundances can potentially be inferred from a detailed measurement of the helioscope spectrum. For each element, a pair of peaks is visible. They correspond to the first two lines of the Lyman series where the lower energy one is the dominant Ly-$\alpha$ line \cite{Redondo:2013wwa}. The Balmer series is only visible in the case of iron but it merges with the Ly-$\alpha$ line of neon. For the purpose of peak detection, a high signal to background ratio is beneficial. Therefore, it is immediately clear that iron will be the best candidate for detection. Depending on the size of the two photon coupling, the Primakoff production of axions provides an additional continuous background. To get an impression of the effect that this would have on different peaks, it is plotted in Fig.~\ref{peak_labels} for an exemplary value of the two photon coupling, $g_{a\gamma}=\SI{e-11}{\per \giga \electronvolt }$. 

\subsection{Measuring the peak strength}
To test whether IAXO would be sensitive enough for measuring solar metal abundances, we employ a simple simulation of the signal. As this is a post-discovery measurement, we take the experimental parameters from the IAXO+ setup~\cite{Armengaud:2019uso}, as briefly summarized in Tab.~\ref{tab:iaxoplus}. A relatively long observation time of five years and a high energy resolution of \SI{10}{\electronvolt} is assumed. We also choose the energy resolution  to be smaller than the width of any of the peaks, which means that high resolution X-ray detectors like metallic magnetic calorimeters would be required~\cite{Kempf2018}.  For simplicity, the axion is assumed to be massless or very light ($m_a\lesssim$ \SI{10}{\milli \electronvolt}) and hence Eq.~\eqref{conversionmassless} can be applied\footnote{The effect of higher masses can be compensated by introducing a buffer gas at the cost of some amount of absorption.}. Putting everything together and using Eq.~\eqref{gammaflux}, a IAXO signal can be generated for arbitrary coupling constants. 

\begin{table}
\centering
	\renewcommand{\arraystretch}{1.2}
\begin{tabular}{|l|l l|c|}
\hline
Magnetic field &$B$ &[T] & 3.5\\
Effective area  &$A$ &[${\rm m}^2$] & 3.9\\
Length &$L$ &[${\rm m}$]  & 22\\
Efficiency &$Q$ & &  0.28\\ 
Observation time &$t$ &[${\rm years}$] & 5 \\\hline
\end{tabular}
\caption{Parameters of the employed IAXO+ setup following Tab.~5 of ~\cite{Armengaud:2019uso}. $Q$ is the combined efficiency of the detector and the optics, which, for simplicity, we take to be energy independent. The background is effectively negligible.}
\label{tab:iaxoplus}
\end{table}

Looking at peaks from one element at a time, the energy region of interest is smaller than $\sim$\SI{1}{\kilo \electronvolt} (100 bins). The expected number of events can be split up into two contributions, one from the metal that we want to detect $\mu_{\mathrm{peak}}$ and one from all other (continuous) processes, which effectively constitute the background $\mu_{\mathrm{back}}$.
We can now ask to what level of precision we can measure $\mu_{\mathrm{peak}}$. As our benchmark scenario, we use the peak strength obtained from OP data and the solar model AGSS09.

It is necessary to calculate the expected relative error on a fine grid in parameter space. To do this in an efficient manner, we use an Asimov data set instead of a Monte Carlo approach (cf.~\cite{Cowan:2010js}). This is defined as the data set where all observables are given by the expectation values that would be obtained from the statistical distribution given the original, ``correct'' input parameter values~\cite{Cowan:2010js}. In our case this corresponds to setting the number of counts in each bin to their respective expectation values. When we apply a likelihood ratio test with an approximate coverage of 95\% to this data, the resulting interval will correspond to the average interval of a large number of measurements. To check that our likelihood ratio test achieves the desired coverage and also that the Asimov approximation is valid, we have performed a full Monte-Carlo simulation with 10000 events for a number of ($g_{a\gamma},g_{ae}$) values. We find good agreement between the two methods.

\begin{figure}
	\centering
	\includegraphics[width=.9\linewidth]{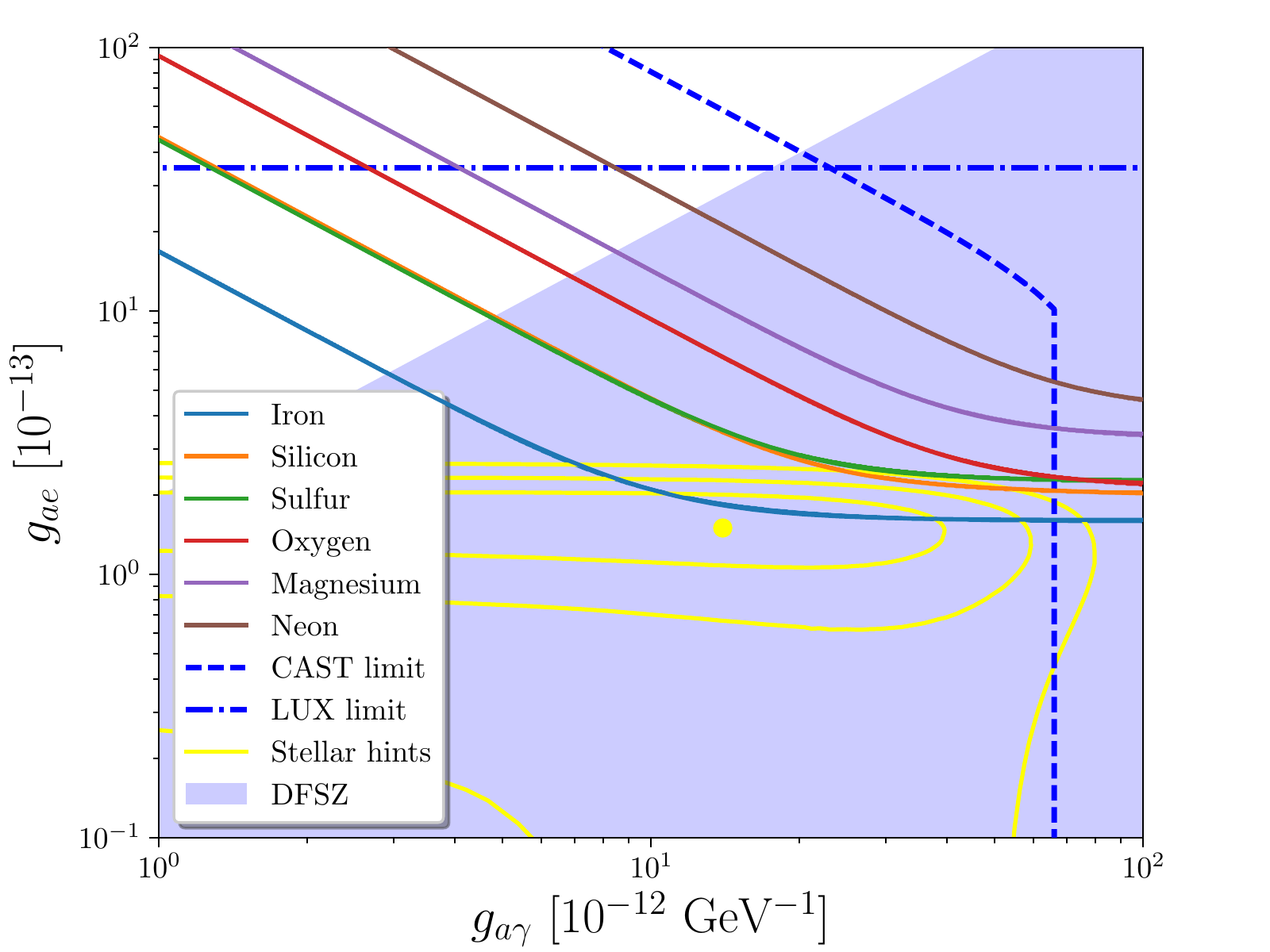}
	\caption{Contours of 20\% relative error in the measurement of the peak strength for several elements using solar axions. Iron is by far the easiest element to detect. Depending on the strength of the Primakoff background at the position of the peak, some elements like oxygen perform better in comparison to others, when the background is Primakoff dominated. Parameter space covered by the DFSZ I and II models is shaded in blue. Above this region, the DFSZ models violate perturbativity \cite{Irastorza:2018dyq,Chen:2013kt,Giannotti:2017hny}. Each point in parameter space has a corresponding model-dependent mass, which we neglect for simplicity. Hints from stellar cooling are taken from~\cite{Giannotti:2017hny} with the 1, 2 and \SI{3}{\sigma} contours and the best fit value shown in yellow. Larger couplings outside the $3\,\sigma$ contour are disfavored by astrophysical observations. Experimental limits from CAST~\cite{Barth:2013sma,Anastassopoulos:2017ftl} (we naively combine the two limits by taking the stronger of the combined electron and photon coupling limit of~\cite{Barth:2013sma} and the pure photon coupling limit of~\cite{Anastassopoulos:2017ftl}) are shown as dashed blue line. Finally, the results of LUX~\cite{Akerib:2017uem}  are depicted by the dashed-dotted blue line.}
	\label{metal-detection}
\end{figure}

Using this technique, we can now calculate the expected accuracy of the peak strength measurement for each point in parameter space.
Fig.~\ref{metal-detection} shows the 20\% accuracy contours for the most relevant elements.
The best results are obtained for iron. In a sizeable part of parameter space that is not excluded by CAST or LUX, the abundance can be measured to at least 20\% accuracy. It even reaches into the region preferred by stellar cooling observations. 
The same method can be applied to all other elements responsible for significant peaks in the axion flux spectrum. 
Elements like oxygen with a low Primakoff background perform better in comparison to other elements when the background is Primakoff dominated (bottom right of Fig.~\ref{metal-detection}).

The asymptotic behaviour of the contours can easily be understood analytically. The number of events in the peaks $\mu_{\mathrm{peak}}\sim g_{a\gamma}^2g_{ae}^2$ have to be comparable to the fluctuation of the background which is proportional to $\sqrt{g_{a\gamma}^4}$ or $\sqrt{g^2_{a\gamma}g^2_{ae}}$ depending on whether the continuous part of the spectrum is dominated by the axion-photon (Primakoff) or the axion-electron coupling. 
In the Primakoff dominated region, we have,
$g_{a\gamma}^2g_{ae}^2\sim \sqrt{g_{a\gamma}^4}$ and consequently $g_{ae} \sim {\mathrm{const}}$.
If the axion-electron coupling is the largest contribution to the continuous part of the spectrum, we have,
$g_{a\gamma}^2g_{ae}^2\sim \sqrt{g_{a\gamma}^2g_{ae}^2}$ and therefore $g_{ae} \sim g_{a\gamma}^{-1}$.
The contours of constant relative errors exhibit this expected asymptotic behaviour.

\subsection{From peak strengths to abundances - Systematic uncertainties}
Naively, the peak strength is directly related to the metal abundance $n_{\mathrm{metal}}$ via,
\begin{equation}
\label{peakstrength}
\mu_{\mathrm{peak}} = g_{a\gamma}^2g_{ae}^2 C_{\rm metal} n_{\mathrm{metal}}.
\end{equation}
We can therefore hope to use the measurements of the peak strength to determine the elemental abundances.

Unfortunately, Eq.~\eqref{peakstrength} suffers from a number of uncertainties. The first ones are the axion couplings. However, since a requirement for the detection of peaks is that the contribution from electron processes is detectable, it is possible to measure $g_{a\gamma}g_{ae}$ with the methods described in~\cite{Jaeckel:2018mbn}. In particular, it will be possible to discriminate the contributions to the signal from axion-electron interactions ($\propto g_{a\gamma}^2g_{ae}^2$) and from Primakoff conversion ($\propto g_{a\gamma}^4$).

\begin{figure}[t!]
	\centering
	\includegraphics[width=.9\linewidth]{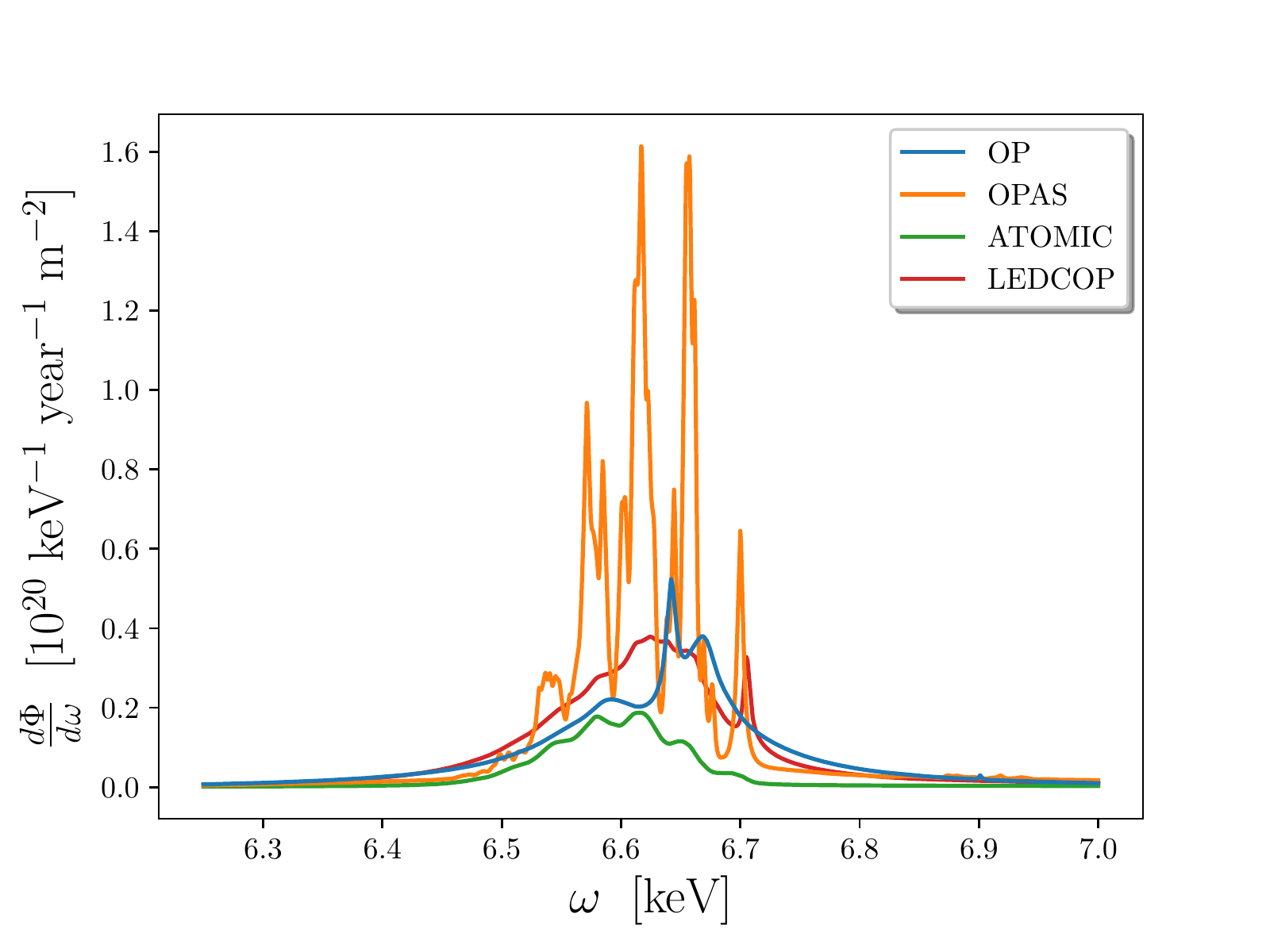}
	\caption{Characteristic axion emission line of iron obtained from the Opacity Project~\cite{Badnell:2004rz,Seaton:2004uz}, OPAS~\cite{OPAS1,OPAS2}, ATOMIC~\cite{Colgan:2016,TOPSpage} and LEDCOP~\cite{Magee:1995,TOPSpage} data. For clarity, we have removed the continuous background.}
	\label{peakspectra}
\end{figure}

The second and indeed most problematic uncertainty is the solar modelling itself, reflected in the constant $C_{\rm metal}$.
The first ingredient is that the metal composition inside the sun may vary as a function of the radius. However, this is most likely not a huge effect as typically more than 90\% of the axion emission originates from radii smaller than $0.3$ times the solar radius (indeed for iron it is less than $0.2$). Within this radius, metal fractions in solar models do not vary dramatically.
A significantly bigger uncertainty arises from the different modelling of the atomic states as reflected in the opacity data. To estimate this, we compare the results obtained with four different opacity sets\footnote{We are indebted to Javier Redondo for this suggestion.}. These are the OP~\cite{Badnell:2004rz,Seaton:2004uz} data used above, the data from OPAS~\cite{OPAS1,OPAS2}\footnote{We would like to thank C.~Blancard and the OPAS collaboration for kindly allowing us to use more detailed data than provided in the publications.} and finally ATOMIC~\cite{Colgan:2016} and LEDCOP\footnote{The publicly available opacity tables generated with ATOMIC and LEDCOP are not regarded as spectroscopically resolved by their authors. Nevertheless, we decided to include them to demonstrate the uncertainties of opacity calculations.}\cite{Magee:1995} which are elemental opacities that can be combined with the TOPS code and are available online at \cite{TOPSpage}. For the example of iron, we show the resulting axion emission (without the continuum contribution) in Fig.~\ref{peakspectra}. We can see that not only the detailed structure is quite different but also the overall emission differs significantly between the data sets. We find similar differences for the other elements that, unfortunately, also depend strongly on the element in question. An overview of the relative peak strength is given in Tab.~\ref{tab:peakstrength}. At present, this uncertainty limits the ability to infer the metal abundances. We note, however, that the OP data, which we used for figure~\ref{metal-detection}, predicts a rather conservative power of the transition peaks. Even the smallest relative power (iron peak in ATOMIC) can be compensated by merely a factor of $0.39^{-\frac{1}{4}}=1.27$ in the coupling constants. Thus, a significant part of parameter space will allow the detection of transition peaks independent of which of the opacity calculations is most accurate.

\begin{table}
\centering
	\renewcommand{\arraystretch}{1.2}
\begin{tabular}{|c|c|c|c|c|c|c|}
\hline
 & Fe & S & Si & Mg & Ne & O\\
 \hline
OPAS/OP  & 1.68& 1.75 & 2.92  & 2.46 & 3.37 & 4.70 \\ 
ATOMIC/OP & 0.39 & 0.83& 1.55  & 1.20 & 0.52 & * \\
LEDCOP/OP & 1.07& 1.05 & 1.51  & 1.52 & 0.90 & * \\ 
\hline
\end{tabular}
\caption{Relative peak strength for different data sets and elements. The $*$ symbols in the oxygen column indicate that for these data sets no clear peak is visible.}
\label{tab:peakstrength}
\end{table}

\section{Conclusions}
If axions are detected in the near future, they can provide a novel probe of the interior of our sun.
We have shown that in suitable regions of the axion parameter space a helioscope like IAXO, equipped with sufficiently good energy resolving detectors, would allow to measure the strength of characteristic emission peaks in the axion spectrum. These peaks are due to the presence of heavier elements, most notably iron, neon and oxygen that are relevant to the solar abundance problem~\cite{Antia:2005mg,Asplund:2009fu}.
As the peak strength is related to the elemental abundance, such a measurement could be an extremely powerful tool in resolving the solar abundance problem. 
Crucially, to realise this potential an improved precision in the modelling of the atomic emission lines inside the plasma is mandatory. Turned around such a measurement could also be used to test this modelling, e.g. by measuring different emission lines of the same element. 

Going beyond that, to measure the total metallicity of the sun would require the independent measurement of the carbon and nitrogen abundances. This is because carbon and nitrogen contribute significantly to the total metallicity ($\sim$ 22\%) but do not cause large peaks in the axion spectrum. Fortunately, carbon and nitrogen abundances could in future be inferred from the solar neutrino flux, as neutrinos from the CNO cycle come within reach of detectors~\cite{Cerdeno:2017xxl}. In this way axion and neutrino experiments could complement each other with neutrinos indicating the light metal abundances and IAXO detecting heavier elements.  

\section*{Acknowledgements}
We would like to thank Javier Redondo for helpful discussions and valuable comments as well as Christophe Blancard and the OPAS team for allowing us to use more detailed data. LT is funded by the Graduiertenkolleg \textit{Particle physics beyond the Standard Model} (GRK 1940). JJ would like to thank the Institute for Particle Physics Phenomenology (IPPP) for splendid hospitality and support by a DIVA fellowship.

\end{document}